\documentclass{article}
\pdfoutput=1 
\usepackage{amsthm,graphicx,amsmath,amssymb,amsfonts,framed,upgreek,physics}   
\usepackage{slashed} 
\usepackage{fullpage,xcolor}
\usepackage{latexsym,slashed,multirow}
\usepackage{graphicx,float}
\usepackage[caption = false]{subfig}

\usepackage{hyperref}
\hypersetup{colorlinks=true, linkcolor=blue, citecolor=red, urlcolor=blue}

\usepackage[nottoc,notlot,notlof]{tocbibind}

\usepackage[utf8]{inputenc}
\usepackage[T1]{fontenc}

\newtheorem{theorem}{Theorem}[section] 
\newtheorem{definition}[theorem]{Definition} 

\def\ds{\partial\!\!\!\slash}

\def\C{{\mathbb{C}}}

\def\I{{\mathbb{I}}}

\def\cinf{{C^\infty(\M)}}

\def\HH{{\cal H}}	
	
\def\M{{\mathcal M}}		
\def\aa{{\mathcal A}}

\def\A{{\mathcal A}}

\def\hh{{\mathcal H}}

\begin{document}

\title{\bf Actions for twisted spectral triple \\[4pt] and
  the transition  from the Euclidean to the Lorentzian}

\author{Agostino Devastato\textsuperscript{**}, \; Manuele Filaci\textsuperscript{*\ddag}, \; 
Pierre Martinetti\textsuperscript{\dag\ddag}, \; Devashish Singh\textsuperscript{\dag}
 \\[8pt] 
 \textsuperscript{**}\emph{Ministero dell'Istruzione, dell'Universit`a e della Ricerca (M.I.U.R.)}, Italy;\\ 
 \textsuperscript{*}\emph{Universit\`a degli Studi di Genova -- Dipartimento di Fisica}, 
 \textsuperscript{\dag}\emph{Dipartimento di Matematica}; \\ 
 \textsuperscript{\ddag}\emph{Istituto Nazionale Fisica Nucleare sezione di Genova,} \\[0pt]  
\emph{via Dodecaneso, 16146 Genova GE Italy.} \\[6pt] 
 \emph{E-mail:} astinodevastato@gmail.com, \,
 manuele.filaci@ge.infn.it,\\ martinetti@dima.unige.it,\, devashish.1612@gmail.com
}
\maketitle

\begin{abstract}
This is a review of recent results regarding the application of
Connes' noncommutative geometry to the Standard Model, and beyond.
By twisting (in the sense of Connes-Moscovici) the spectral triple of
the Standard Model, one does not only get an extra scalar field which
stabilises the electroweak vacuum, but also an unexpected $1$-form
field.
By computing the fermionic action, we show how this field induces a
transition from the Euclidean to the Lorentzian signature. Hints on a
twisted version of the spectral action are also briefly mentioned.
\end{abstract}

\section{Introduction}	

Noncommutative geometry ``a la  Connes'' \cite{Connes:1994kx}
allows to obtain the Lagrangian of the Standard Model of elementary
particles - including the Higgs sector - minimally coupled with Einstein-Hilbert action (in Euclidean
signature) from geometrical principles. In addition, it offers some guidelines to go beyond
the Standard Model by playing with the mathematical rules
of the game (for a recent review,~see~\cite{Chamseddine:2019aa}). 

Early attempts ``beyond the SM'' were considering new fermions (see e.g. \cite{Stephan:2009fk}
and other papers of the same author).
One may also relax one of the axioms of noncommutative geometry,  the
first-order condition discussed below
\cite{Chamseddine:2013uq};
or modify another axiom regarding the real structure (also discussed
below) \cite{T.-Brzezinski:2016aa,Brzezinski:2018aa}.
Other proposals are based on some Clifford bundle structure \cite{Dabrowski:2017aa},
or non-associativity \cite{Boyle:2014aa}. 
Here we focus on a model consisting in twisting the original noncommutative geometry
  \cite{Devastato:2015aa,buckley,Devastato:2014fk}.

From the examples listed above, all but the first are minimal extensions of the
Standard Model: they allow to produce the kind of extra scalar field $\sigma$ suggested
by particle physicists to stabilise the electroweak-vacuum (which
also permits to make the calculation of the Higgs mass in
noncommutative geometry compatible with its experimental value), without adding new fermions.

By using twisted noncommutative geometry, one gets in addition
  an supplementary piece, namely a $1$-form field, which surprisingly turns out to
  be related to the transition from Euclidean to Lorentzian signature.

We give an overview of these results below, beginning in \S
\ref{sec:sm-ncg}  with a recalling on the Standard Model in
noncommutative geometry. Then we summarise in \S \ref{sec:twisted} how
to apply a twist to the geometry, and why this is related to a
transition from the Euclidean to the Lorentzian. In \S
\ref{sec:actions} we show how this transition is actually realised at
the level of the fermionic action. We also stress some projects
regarding the spectral (i.e. bosonic) action.

 \section{The Standard Model in noncommutative geometry}
\label{sec:sm-ncg}

\subsection{Spectral triple}
\begin{definition}\cite{Connes:1996fu}
  A spectral triple consists of an involutive algebra $\A$ acting on a
  Hilbert space $\HH$, together with selfadjoint operator $D$ on $\HH$ such
  that the commutator $[D,a]$ is bounded for any $a\in\A$.
It is graded if, in addition, there exists a selfadjoint operator
$\Gamma$ which squares to $\I$, and such that 
\begin{equation}
\left\{\Gamma, D\right\} =0, \quad [\Gamma, a] = 0 \quad \forall a\in \A.
\label{eq:2}
\end{equation}
\end{definition}
Any closed Riemannian (spin) manifold $\M$ defines a spectral
  triple
  \begin{equation}
\label{eq:009}
\cinf\,,\quad  L^2(\M, S)\,,\quad \ds=-i\gamma^\mu \nabla_\mu
\end{equation}
where $\cinf$ is the algebra of smooth functions on $\M$, acting by
multiplication on the Hilbert space $L^2(\M,S)$ of square integrable
spinors, and $\ds$ 
is the Dirac operator, with $\nabla_\mu=\partial_\mu + \omega_\mu$ the
covariant derivative associated with the spin connection
$\omega_\mu$. For an even dimensional manifold, the spectral triple is
graded with grading  the product of the Dirac matrices,  that
is $\gamma^5$ for a four dimensional manifold. 

A supplementary structure that plays an important role in the
construction of physical models is the \emph{real structure} \cite{Connes:1995kx}. The
latter consists of an antilinear operator $J$ such that 
\begin{equation}
  J^2=\epsilon\I,\; JD=\epsilon' DJ,\; J\Gamma=\epsilon''
  \Gamma J
\end{equation}
 where $\epsilon, \epsilon', \epsilon''=\pm 1$
 define the
 so-called $KO$-dimension $k\in\left[0,7\right]$ of the spectral triple.
In addition, the operator  $J$ implements a map
$a\to a^\circ :=  Ja^*J^{-1}$ from $\A$ to the opposite
algebra $\A^\circ$ (the same object of $\A$ as a vector space, but
with opposite product: $a^\circ b^\circ = (ba)^\circ$). This allows to
define a right action of $\A$ on $\HH$,
\begin{equation}
  \psi a := a^\circ \psi, 
\end{equation}
which is asked to commute with the left action (\emph{the order zero condition})
\begin{equation}
[a, Jb^*J^{-1}] = 0 \quad\quad\forall a,b \in \A.\label{eq:6}
\end{equation}
 
For the spectral triple (\ref{eq:13}), the real structure is
${\cal J}=i\gamma^0\gamma^2cc$ where $cc$ denotes the complex
conjugation. This coincides with the charge conjugation operator of
quantum field theory.
 
Finally, one requires that the following \emph{first order condition} holds
\begin{equation}
[[D,a], Jb^*J^{-1}] = 0\quad\quad\forall a,b \in \A
\label{eq:7}
\end{equation}
which is an algebraic formulation of $D$ being
a first-order differential operator.
\smallskip

With other extra-conditions, spectral triples provide a
\emph{spectral characterization of manifolds} \cite{connesreconstruct}. Namely, any closed
Riemannian (spin) manifold defines a spectral triple (\ref{eq:13});
conversely, given a spectral  triple $(\A, \HH, D)$ with $\A$ commutative
  and unital, then there exists a Riemannian manifold $\M$ such that $\A=\cinf$.

This motivates the definition of a \emph{noncommutative geometry} as a spectral triple
$(\A, \HH, D)$ where $\A$ is non-necessarily commutative.

\subsection{Gauge theory}

A gauge theory (on a four dimensional manifold $\M$) is described by
\cite{Connes:1996fu, Connes:2008kx} the product, in the sense of spectral
triple, of a manifold (\ref{eq:009}) by a finite-dimensional spectral triple
\begin{equation}
  \label{eq:12}
  \A_F, \;\HH_F,\; D_F
\end{equation}
that encodes the gauge degrees of freedom. The product triple is
\begin{equation}
  \label{eq:13}
  \A=\cinf\otimes \A_F,\quad\HH= L^2(\M,S)\otimes \HH_F, \quad D=\ds\otimes
  \I_F + \gamma^5 \otimes D_F
\end{equation}
where $\I_F$ is the identity operator on $\HH_F$.

The connection  $1$-forms, generalised to the noncommutative setting, are elements of 
\begin{equation}
\Omega^1_D(\A):=\left\{ \Sigma_i  a_i[D,b^\circ_i]\right\}, \quad a_i, b_i\in\A,
\label{eq:9}
\end{equation}
and the associated covariant Dirac operator is 
\begin{equation}
D_A= D+A + J\, A \,J^{-1} \quad\text{ with } A\in \Omega_D^1(\A).\label{eq:16}
\end{equation}

A gauge transformation is implemented by the conjugate action of
\begin{equation}
\text{Ad}(u): \psi \to u\psi u^* = u (u^*)^\circ \psi = u J
uJ^{-1}\psi,
\label{eq:21}
\end{equation}
for $u$ is a unitary element of $\A$. Namely, a gauge transformation
maps the covariant operator $D_A$ to 
\begin{equation}
\text{Ad}(u)\, D_A \, \text{Ad}(u)^{-1} = D_{A^u}\label{eq:22}
\end{equation}
where $A^u$ is the gauge transformed of the potential $A$, given by 
\begin{equation}
A^u:= u[D, u^*] + uAu^*.\label{eq:23}
\end{equation}

\subsection{The Standard Model}

The finite dimensional spectral triple that describes the Standard
Model is \cite{Chamseddine:2007oz}
\begin{equation}
  \A_{F} = \C \oplus {\mathbb H} \oplus M_3(\C), \quad \;
  \hh_{F} = \C^{32 =  2 \times 2 \times 8} ,\quad D= D_0 + D_R
\end{equation}
where $\mathbb H$ is the algebra of quaternions. The algebra is such that
its group of unitary element gives back the gauge group of the
standard model, and $32$ is the number of fermions per generation ($6$
coloured quarks and two leptons, that exists in two chiralities,
together with their antiparticles). The operator $D$ is a $32\times
32$ matrix, which divides into a block diagonal part
$D_0$ which contains the Yukawa couplings of  the
  electron, the quarks up and down,
  and the (Dirac) mass of the electronic neutrino; and an off-diagonal
  part $D_R$ which contains only one non-zero
  entry $k_R$ (Majorana mass of the electronic neutrino). The
  structure is then repeated for the other two generations of fermions.

A generalised $1$-form (\ref{eq:9}) then divides into two pieces,
  \begin{equation}
    A = \gamma^5\otimes H -i\sum_\mu \gamma^\mu \otimes A_\mu,
  \end{equation}
where  $H$ is a scalar field on $\M$ with value in
 $\aa_F $, that identifies with the Higgs field, while $A_\mu$ is a
 $1$-form field with value in $Lie(U(\aa_F))$, whose components give
 the gauge fields of the standard model.

The fermionic action of the Standard Model is retrieved as
\begin{equation}
  \label{eq:14}
S^f(D_A)= \frak A_{D_A}(\tilde \xi, \tilde \xi)
\end{equation}
with $\tilde \xi$ the Grassman variables associated to a $+1$
eigenvector of the grading operator, 
and
\begin{equation}
\label{eq:fermac}
\frak A_{D_A}(\xi, \xi') =\langle J\xi, D_A\xi'\rangle.
\end{equation}
is the bilinear form defined by the covariant Dirac operator and the
real structure.  The asymptotic expansion $\Lambda\to\infty$ of the  \emph{spectral action}
  \begin{equation}
  \text{Tr}\; f\left(\frac{D_A^2}{\Lambda^2}\right)
\label{eq:10}
\end{equation}
($f$  being a smooth approximation of the characteristic function of
$[0,1]$)  yields the  bosonic Lagrangian of the standard model
 coupled with the
 Einstein-Hilbert action in Euclidean signature.

 The spectral action provides initial conditions at a putative unification
 scale. 
Physical predictions are obtained by running down
the parameters of the theory under the renormalisation group equation.
Assuming there is no new physics between the unification scale and our
scale, one finds a mass of the Higgs boson around  $170 \text{ GeV}$, which is not in agreement with the experimental
 value $m_H=125,1\text{GeV}$.

 But
it was well known in particle physics that for a Higgs boson with mass $m_H\leq 130 \,\text{Gev}$, the quartic
coupling of the Higgs field becomes negative at high energy, meaning the
electroweak vacuum is meta-stable rather than
stable \cite{Degrassi:2012fk, near-critic}.
Such instability can be cured by a
new scalar field $\sigma$, that couples to the Higgs in a suitable
way \cite{Elias-Miro:2012ys}. 

In the spectral triple of the Standard Model, such a field $\sigma$
 is obtained by turning into a field  the
 neutrino Majorana mass~$k_R$ which appears in the off-diagonal
part $D_R$ of the finite dimensional Dirac operator $D_F$ \cite{Chamseddine:2012fk}:
$$\ {k_R\to k_R\sigma}.$$
In addition, by altering the~running of the parameters under the equations of the group
of renormalisation,  $\sigma$ makes the computation of $m_H$ compatible
with $125$~Gev. 
 
The point is that the field $\sigma$ cannot be obtained on the same
footing as the Higgs, that is as a component
of a generalised $1$-form \eqref{eq:9}, for
\begin{equation*}
  \label{eq:5}
    [\gamma^5\otimes D_R, a]= 0 \quad\quad\forall a, b\in \A=\cinf\otimes \A_{F}.
\end{equation*}
This motivates to modify the spectral triple of the Standard
Model. Several ways have been explored, some of them listed in the
introduction. Here we follow the path consisting in twisting the
spectral triple. This is a way to implement on a solid mathematical
ground the idea of grand-symmetry, first  introduced in \cite{Devastato:2013fk}.

\section{Twisted spectral triples and Lorentz signature}
\label{sec:twisted}

\subsection{Minimal twist of the Standard Model}

 Given a triple $(\A, {\cal H}, D)$, instead of asking the commutators
 $[D,a]$ to be bounded, one asks the boundedness of the
twisted commutators 
\begin{equation}
[D, a]_\rho := Da - \rho(a) D \quad \quad\text{ for some fixed} \quad \rho\in \text{Aut}(\A).
\label{eq:11}
\end{equation}
Such a variation of the original definition of the spectral triples
were introduced in \cite{Connes:1938fk} with purely mathematical
motivations. Later it was realised that twisted spectral triples
provide a way to generate the field $\sigma$ required to stabilise the
electroweak vacuum and fit the calculation of the Higgs mass \cite{buckley}. 

The idea is to introduce a twisting automorphism $\rho$ in the spectral
triple of the Standard Model, with minimal changes. In particular, we keep the Hilbert space and the
Dirac operator unchanged, since they encode the
fermionic content of the theory, and there are so far no indications of
new fermions beyond those of the Standard Model. These requirements
make necessary  to double the algebra \cite{Lett.}. Namely, one
considers the triple
\begin{equation}
\A=(\cinf\otimes \A_F)\otimes \ {\C^2},\quad
 {\HH} = L^2(\M, S)\otimes \HH_F,  \quad \ {D}=\ds\otimes
\I_{32}+\gamma^5\otimes D_{\text{F}}\label{eq:15}
\end{equation}
with automorphism $\rho$ the flip
$$\rho((f,g)\otimes m)  = (g,f)\otimes m \quad\quad  f,g \in\cinf,
m\in \A_F.$$ 

Instead of (\ref{eq:9}), one considers  a
twisted generalised form
\begin{equation}
  \label{eq:5}
  A_\rho=  \Sigma_i  a_i [D, b_i^\circ]_\rho,  \quad a_i, b_i\in \A,
\end{equation}
and the associated twisted-covariant Dirac operator 
\begin{equation}
D_{A_\rho}:= D+ A_\rho+J \, A_\rho \, J^{-1}.
\label{eq:8}
\end{equation}
One generates in this way the extra-scalar field $\sigma$, since the twisted
commutator
$ [\gamma^5\otimes D_R, a]_\rho$  is no longer zero, and yields precisely the kind of scalar field
 $\sigma$ discussed above.

But there is also an
unexpected guest, namely a $1$-form field $f_\mu dx^\mu$, coming from the
twisted commutator of the free part $\ds\otimes \I_F$. 

\subsection{Lorentzian inner product from twist}

A  gauge transformation for a twisted spectral triple is given by the
twisted conjugate action of the operator $U=\text{Ad}(u)$ in (\ref{eq:21}), that is
\begin{equation}
\ { \rho(U) \, D_{A_\rho} \, U^{-1}=D_{A_\rho^u} }\label{eq:19}
\end{equation}
where $ {\rho(U) =\rho(u)J\rho(u)J^{-1}}$ with $u$ a unitary of
$\A$  and \cite{Landi:2017aa}
\begin{equation}
\ {A_\rho^u:= \rho(u) [D,u^*]_\rho + \rho(u)Au^*}.\label{eq:20}
\end{equation}
This is a twisted version of the law of transformation of the gauge
potential (\ref{eq:23}). 

The main difference with the usual gauge transformation (\ref{eq:22}) 
is that the latter preserves the selfadjointness of the operator
$D_A$, whereas (\ref{eq:19}) does not preserve the selfadjointness of
$D_{A_\rho}$. However, it preserves the adjointness with respect to the inner product
induced by the twist, which is defined as follows.

\begin{definition}
Let $\rho$ be an automorphism of the algebra ${\cal B}(\HH)$ of bounded operators on an
Hilbert space $\HH$. A  $\rho$-twisted inner product
   $\langle\cdot ,\, \cdot \rangle_\rho$ is an inner product on $\HH$ such
  that
 $$ \langle\Psi,\mathcal{O}\Phi\rangle_{\rho}=\langle\rho(\mathcal{O})^\dag\Psi,\Phi\rangle_{\rho}\qquad
 \forall {\cal O}\in{\cal B}(\HH),\; \Psi,\,\Phi\in\HH,$$
where $^\dag$ is the adjoint with respect to the initial inner
product. We denote the $\rho$-adjoint of ${\cal O}$ as
${\cal O}^+:= \rho({\cal O})^\dag.$
\end{definition}

 If $\rho$ an inner automorphism of ${\cal B}(\HH)$, that is 
$\rho({\cal O}) = R{\cal O} R^\dag$ 
for a unitary operator $R$ on $\HH$, then a natural $\rho$-product is 
$$\ {\langle\Psi,\Phi\rangle_\rho=\langle\Psi, R\Phi\rangle}.
$$

In the twisted spectral triple of the Standard Model, the flip $\rho$ is an inner automorphism of
${\cal B}(L^2(\M, S))$, with 
$R=\gamma^0$.
Thus the $\rho$-twisted inner product is nothing but the  Krein product for the
  space of spinors on a \emph{Lorentzian} manifold.
Furthermore, extending $\rho$ to the whole of ${\cal B}(L^2(\M, S))$,
one finds
$$\rho(\gamma^0)=\gamma^0, \quad \rho(\gamma^j)=-\gamma^j \quad
\text{for} \quad j=1,2,3.$$The flip $\rho$ is the  in some sense the
square of the Wick rotation \cite{DAndrea:2016aa}
$$W(\gamma^0)=\gamma^0, \quad W(\gamma^j)=i\gamma^j.$$
that is
$\rho=W^2.
$

This suggests that the twisting procedure has something to do with 
a transition from the Riemannian to the Lorentzian signatures. This is
confirmed by studying the fermionic action for a twisted
spectral triple. 

\section{Actions for twisted spectral triples}
\label{sec:actions}

\subsection{Twisted fermionic action}

The fermionic action $S^f_\rho$ for a twisted spectral triple is defined \cite{Devastato:2018aa}
substituting
the inner product in (\ref{eq:fermac}) with the twisted
$\rho$-product, and considering the twisted covariant operator
$D_{A_\rho}$ instead of $D_A$. Also, one does not restrict to an
eigenspace of the grading operator, but consider instead an eigenvector of
the unitary $R$ that implements the twist. This is required to
guarantee that the fermionic action is antisymmetric as a bilinear
form, allowing thus the switch to Grassmann variables.

This has important consequences, most
easily seen in the simplest example of the  minimal twist
 of  a manifold (of
even dimension $2m$):
\begin{equation*}
   \label{eq:184}
\A= \cinf\otimes\C^2, \quad \HH= L^2(\M,S), \quad  D=\ds;  \quad \rho
 \end{equation*}
where the representation $\pi$ of $\A$ on $\HH$ is
 \begin{equation*}
   \label{eq:4}
   \pi(f,g)=\left(\begin{array}{cc} f\,\I_{2^{m-1}}& 0 \\  0& g\I_{2^{m-1}}\end{array}\right),
 \end{equation*}
 and $\rho$ is the flip
\begin{equation}
\rho(f,g) = (g,f) \quad\quad \forall 
(f,g)\in \A\simeq \cinf\oplus\cinf.
\label{eq:187}
\end{equation}

A twisted generalised $1$-form is parametrised by a $1$-form field
$f_\mu$ (there is no scalar field $\sigma$). The twisted  fermionic
action, in dimension $4$ has been computed in
\cite{Martinetti:2019aa}. One finds
\begin{equation}
S^f(\partial_{\rho}) = 2\int_\M d\mu\;  \bar{
\tilde \zeta}^\dagger\sigma_2\,(if_0\I_2-\sum_{j=1}^3\sigma_j\partial_j)\, \tilde
\zeta \quad \text{ where }\quad 
\xi=\left(\begin{array}{c}\zeta\\ \zeta\end{array}\right)\in
\HH_R,\label{eq:18}
\end{equation}
with ${\cal H}_R$ the $+1$ eigenspace of the unitary $R$. 
The striking point is the disappearance of the derivative $\partial_0$, substituted with the component $f_0$ of
the twisted fluctuation.
It reminds the Weyl Lagrangian
$\psi_l^\dag\,
  \tilde\sigma_M^\mu\, \partial_\mu \psi_l$ where $\tilde\sigma_M^\mu :=\left\{\mathbb I_2, -\Sigma_{j=1}^3\sigma_j\right\}$.
Actually, 
it is tempting to  identify $\tilde \zeta$ with $\psi_l$, then to assume
$$\partial_0 \psi_l = if_0 \tilde \zeta,$$ that is 
$$\tilde
\zeta (x_0, x_j)=\psi_l(x_0,  x_j)= 
e^{itf_0}\psi_l(x_j).$$
But then $\bar{\tilde  \zeta}^\dag  \sigma^2$ has no reason to
identify with $i\psi_l^\dag$.
In other terms, there are not enough degrees of freedom to identify
(\ref{eq:18}) with the Weyl Lagrangian.

This is cured by considering a double manifold, that is  
\begin{equation}
  \label{eq:25}
 {\cal A} = \left(C^\infty({\cal M}) \otimes \mathbb{C}^2\right) \,
\otimes \,\C^2, \quad
	{\cal H} = L^2({\cal M,S}) \otimes \mathbb{C}^2, \quad
	D = \eth \otimes \mathbb{I}_2
\end{equation}
with representation 
\begin{equation}
	\pi({\small a=(f,g)},a'=(f', g')) =
	\left(\begin{array}{cccc}
		f\mathbb{I}_2 & 0 & 0 & 0 \\ 0 & f'\mathbb{I}_2 & 0 & 0 \\ 
		0 & 0 & g'\mathbb{I}_2 & 0 \\ 0 & 0 & 0 & g\mathbb{I}_2
	\end{array}\right).
\end{equation}
The minimal twist is then given by $\A\otimes \C^2$ acting on the same
Hilbert space, with the same Dirac operator, and the automorphism is
the flip. Then the twisted fermionic action gives back the Weyl
equation in Lorentzian signature.

A similar result holds for the minimal twist of the spectral triple of
electrodynamics \cite{Dungen:2011fk} in Euclidean signature. The
twisted fermionic action yields the Dirac equation in
\emph{Lorentzian signature} (and in the temporal gauge of Weyl).

As a conclusion, the component $f_0$ of the $1$-form field that
parametrises a general twisted $1$-form (\ref{eq:5}) gets interpreted
as the energy of a plane wave solution of the Weyl/Dirac equation in Lorentzian
signature, even though one started with a Riemannian manifold. After a
Lorentz transformation, the other components $f_i$, $i=1,2,3$ get interpreted
as spatial momenta (see \cite{Martinetti:2019aa} for details).
 
\subsection{Spectral action for twisted spectral triple}
 \label{sec:spectral}

To adapt the spectral action (\ref{eq:10}) to the twisted case,
there are several options that are still ``work in progress''.
First of all, 
since the selfadjointness of the twisted covariant Dirac operator  $D_{A_\rho}$ is not preserved
by a twisted gauge transformation (\ref{eq:19}),  $(D_{A_\rho^u})^2$
may not remain positive (nor selfadjoint. not even normal), so that
there is no guaranty to make sense of $f(D_{A_\rho^u})$ thanks to the spectral
theorem. This difficulty can be overcome by considering $(D_{A_\rho})^\dagger (D_{A_\rho})$
instead of $(D_{A_\rho})^2$, as was done in \cite{buckley}. As noticed in
\cite{Devastato:2018aa}, under a twisted gauge transformation 
$(D_{A_\rho})^\dagger  D_{A_\rho}$ is mapped to $UD_{A_\rho}^\dagger
D_{A_\rho}U^\dag$, which has the same trace as
$D_{A_\rho}D_{A_\rho}^\dag$. Hence the action
\begin{equation}
  \label{eq:27}
  \text{Tr} f\left(\frac{(D_{A_\rho})^\dagger D_{A_\rho}}{\Lambda^2}\right)
\end{equation}
is well defined and gauge invariant.

Alternatively, one may use the twisted $\rho$-product of definition 2,
and consider the trace of
$(D_{A_\rho})^+D_{A_\rho}$. Although the gauge invariance is not
obvious and, for the same reasons explained above, the cut-off by
$f(\frac{\cdot}{\Lambda})$
is not guaranteed by the spectral theorem, it is intriguing that in the case of the minimal twist of a
Riemannian manifold, one has \cite{Manuel-Filaci:2020aa}
\begin{equation}
  \label{eq:28}
  D^+D=-(\partial_0)^2 + \Sigma_j (\partial_j)^2 + 2i\gamma^0\gamma^j\partial_j\partial_j,
\end{equation}
that is the sum of the squared of the free Dirac operator on
\emph{Minkowski space} $\ds_M$ with a correction term with vanishing trace.
This tends to confirm the idea
that a transition from the Euclidean to the Lorentzian does occur at the
level of the action, if this is not at the level of the $\gamma$
matrices. To deal with the cut-off, one should use some technics of
algebraic quantum field theory. In particular,  one may select positive
frequencies using another state than the trace truncated by the energy
cut-off $\Lambda$. This will be
investigated in some future works.

A third option is to consider from the start a $\rho$-adjoint Dirac
operator, for example the free Dirac operator in Minkowski space
$\ds_M$.  Then \cite{Devastato:2018aa} 
  $\ds_M^\dagger \ds_M$ is (up to a sign), the Laplacian in Euclidean
  signature. The transition is then from the Lorentzian to the
  Riemannian. A similar calculation  with a $\rho$-adjoint twisted covariant Dirac operator
 yields results similar to those of
  \cite{buckley}. This will be explained in some future work.  

More generally, this last example questions the definition of twisted
spectral triple: would it make sense to impose the Dirac operator to
be $\rho$-adjoint with respect to the twisted product rather than imposing the selfadjointness with
respect to the initial Hilbert product ?

\bibliographystyle{abbrv}
  \bibliography{/Users/pierre/physique/articles/bibdesk/biblio}

\end{document}